\documentclass[aps,prb,twocolumn,superscriptaddress]{revtex4-1}
\usepackage{graphicx}
\usepackage{amssymb}
\usepackage{amsmath}
\usepackage{amsfonts}
\usepackage{color}

\begin{document}

\title{Triple helix vs. skyrmion lattice in two-dimensional non-centrosymmetric magnets} 
\author{V.~E. Timofeev}
\email{timofeeviktor@gmail.com}
\affiliation{NRC ``Kurchatov Institute", Petersburg Nuclear Physics Institute, Gatchina
188300, Russia}
\affiliation{St. Petersburg Electrotechnical University ``LETI'', 197376 St. Petersburg, Russia}
\author{A.~O. Sorokin}
\affiliation{NRC ``Kurchatov Institute", Petersburg Nuclear Physics Institute, Gatchina
188300, Russia} 
\author{D.~N. Aristov}
\affiliation{NRC ``Kurchatov Institute", Petersburg Nuclear Physics Institute, Gatchina
188300, Russia}
\affiliation{St.Petersburg State University, 7/9 Universitetskaya nab., 199034
St.~Petersburg, Russia} 
 
\begin{abstract}
It is commonly assumed that a lattice of skyrmions, emerging in two-dimensional non-centrosymmetric magnets in external magnetic fields, can be represented as a sum of three magnetic helices. In order to test this assumption we compare two approaches to a description of regular skyrmion structure. We construct (i) a lattice of Belavin-Polyakov-like skyrmions within the stereographic projection method, and (ii) a deformed triple helix  defined with the use of  elliptic functions. The estimates for the energy density and magnetic profiles show that these two ansatzes are nearly identical at zero temperature for  intermediate magnetic fields. However at higher magnetic fields, near the transition to topologically trivial uniform phase,   the stereographic projection method is preferable, particularly,  for the description of  disordered skyrmion liquid phase. 
We suggest to explore the intensities of the secondary Bragg peaks to obtain the additional information about the magnetic profile of individual skyrmions. We estimate these intensities to be several percents of the main Bragg peak at high magnetic fields. 

\end{abstract}

\maketitle

\section{Introduction}

Topologically protected states of matter attract the attention of researchers from various fields of science. One of the well-known example of topologically protected objects are skyrmions. Despite the fact that the first appearance of skyrmions is associated with particle physics \cite{Skyrme1962}, the study of magnetic skyrmions has become a rapidly developing field of condensed matter physics over the last decade \cite{Nagaosa2013,garst2017collective}. Most discussed magnetic skyrmions are nanoscale vortex-like configurations. A relatively small size of skyrmions makes them promising objects for the developing of new types of data storage devices \cite{Koshibae_2015,Tomasello2014}. According to the Hobart-Derrick theorem \cite{derrick1964},  topological arguments alone are not enough to stabilize skyrmions, while additional conditions are needed to fix a skyrmion size. Usually, a single skyrmion or a unordered set of skyrmions can be stabilized in a finite sample: a disc\cite{metlov13vortex} or a track (nanoribbon)\cite{Fert2013}. In this case, the stability of skyrmions is provided by the dipole-dipole interaction and surface effects.   For an infinite system the stabilization of skyrmions is achieved in  non-centrosymmetric magnets, where the combination of the Dzyaloshinskii-Moriya interaction (DMI)\cite{dzyaloshinskii1964theory} and an applied magnetic field lead to an existence of long-period modulated magnetic phases, so that single skyrmions appear as elements of a so-called skyrmion crystal (SkX)\cite{bogdanov1989thermodynamically}.  Probably the best studied class of non-centrosymmetric magnets is $B20$ compounds, including MnSi, etc.\cite{muhlbauer2009skyrmion} 

Experimental investigations of such compounds show that the skyrmion phase in the bulk  (also called as A-phase) exists at finite temperatures, slightly below the critical one, $T_{c}$. Thermal fluctuations are expected to play a crucial role in the stability of A-phase\cite{muhlbauer2009skyrmion}. This phase is observed at moderate magnetic fields, with its phase boundary far away from the critical (saturation) field. The intensity maps of neutron scattering experiments show a hexagonal pattern of Bragg peaks in the A-phase region. It allows to interpret the A-phase spin configuration in two ways: either as a hexagonal skyrmion superlattice or as a sum of three simple helices with wave-vectors directed at an angle of 120 degrees relative to each other \cite{muhlbauer2009skyrmion}. These two descriptions are not equivalent and may be distinguished in experiments, but the corresponding difference may be hidden by the experimental specifics and thermal modulation of the local magnetization \cite{adams11}. The latter reason makes thin films investigations more preferable, where the A-phase is more stable and exists \cite{Yu2010b} at $T\approx0$. 

It is known that the correspondence between long-period modulated phases (like a helix) and phases with a finite soliton density may be exact. One  such example happens in one spatial dimension, where skyrmions are kinks in the sine-Gordon model \cite{Skyrme-Gordon62,Perring1962}. A one-dimensional magnet with uniaxial anisotropy, DMI and an external field is described by the sine-Gordon model with the Lifshitz invariants. This model has been exactly solved by Dzyaloshinskii as a modified helical configuration in terms of Jacobi elliptic functions \cite{dzyaloshinskii1964}. As an alternative (dual) description of this solution, one can consider a lattice of kinks \cite{Borisov1988,Aristov02,Borisov09}.

The two-dimensional case is more difficult for modeling. Due to non-linearity, the triple helix anzatz as a sum of three helices is not an exact solution for the ground state at $T=0$. Moreover, one can propose several ways to construct a "triple helix" configuration. The simplest way, usually found in literature   (see, e.g. \cite{muhlbauer2009skyrmion, Tatara2014}) is a sum of ordinary (non-modified) helices  \cite{kaplan1961}.

Recently we showed \cite{Timofeev2019207} that the stereographic projection method provides very good estimate of the ground state energy, and the shape of the individual skyrmions  remains nearly invariant under pressure from its neighbors.  The advantage of the latter method is its flexibility what concerns the the positions and sizes of individual skyrmions. One can particularly employ this way of description for the skyrmion liquid state reported previously in \cite{seshadri1991hexatic} and in \cite{huang2020melting} at some magnetic fields. 

In this paper, we examine different descriptions of skyrmion lattice state in two dimensions at zero temperature. In Section \ref{sec:SkX}, we describe the stereographic approach for the skyrmion crystal construction. In Section \ref{sec:SingleH}, we remind a general form of the magnetic helix for systems with DMI and magnetic field in terms of the additional elliptic parameter \cite{izyumov1984modulated}. With this generalization, we construct  the triple helix ansatz in section \ref{sec:TripleH} at $T=0$ with normalization conditions for the local magnetization. In Section  \ref{sec:comparison} we  compare the  modeling by Skyrmion crystal and triple helix with respect to density of classical energy, the period of the spatial modulation, and intensities of higher-order Bragg peaks.  Our final remarks are presented in Section \ref{sec:conclu}.

\section{Skyrmion crystal} \label{sec:SkX}

We consider the two-dimensional system characterized by magnetization   ${\bf S}({\bf r})$. At zero temperature the magnetization is saturated and can be normalized, ${\bf S}^2=1$. 
The classical energy density in the standard model of chiral magnets is 
\begin{equation}
\mathcal{E} =     \tfrac{1}{2} C \partial_{\mu}S^{i}\partial_{\mu}S^{i} - D
\epsilon_{\mu ij} S^{i}\partial_{\mu}S^{j}  + B(1 - S^{3})  \,,
\label{energy}
\end{equation}
where $\mu=1,2$ and $i=1,2,3$. The first term corresponds to the ferromagnetic exchange, the second one is DMI, and the last one is the Zeeman energy related to an external magnetic field perpendicular to the plane. The main spatial scale in this model is defined by $L=C/D$ and the energy scale is $D^{2}/C$. 
After appropriate rescaling Eq.\  \eqref{energy} reads
\begin{equation}
\mathcal{E} =     \tfrac{1}{2}\partial_{\mu}S^{i}\partial_{\mu}S^{i} -
\epsilon_{\mu ij} S^{i}\partial_{\mu}S^{j}  + b(1 - S^{3})  \,,
\label{dimenergy}
\end{equation}
with the dimensionless magnetic field $b=CB/D^2$.

A single skyrmion is an axially symmetric solution with a unit topological charge. 
Multi-skyrmion configurations can be  described in the stereographic projection approach  
\cite{Timofeev2019207}, which is a convenient way to take into account the interaction between skyrmions and construct fully periodic configuration of SkX. In this section we sketch the main idea  of such consideration.

For the normalized solution one can write
\begin{equation}
\begin{aligned}
S^{1} + i S^{2} & = \frac{2f(z,\bar{z})}{1 + f(z,\bar{z})\bar{f}(z,\bar{z})}   \,, \\
S^{3} &= \frac{1 - f(z,\bar{z})\bar{f}(z,\bar{z})}{1 + f(z,\bar{z})\bar{f}(z,\bar{z})}  \,, \\
\end{aligned}
\label{stereodef}
\end{equation}
where  $f(z,\bar{z})$ is a complex-valued function of $z=x+iy$ and $\bar{z}=x-iy$. It was noticed early on  \cite{Belavin1975} that every holomorphic or antiholomorphic function is a solution of the model without both DMI and an external magnetic field. One can check in the latter case that  one skyrmion corresponds to $f=z_0/\bar{z}$, and that $N$- skyrmion solutions are given by   $f=\sum_{j=1}^{N}z_0^{j}/(\bar{z}-\bar{z_j})$ ; here 
 $z_0^{j}$ define radii and orientation of individual skyrmions. 

When we discuss single skyrmion solution, addition of DMI and external  field   may lead to continuous transformation of the Belavin -- Polyakov (BP) solution, without changing the character of singularities. Our ansatz for the single skyrmion solution is given by:
\begin{equation}
\begin{aligned}
f(z,\bar{z}) = \frac{e^{i \alpha} \kappa (z\bar{z})}{\bar{z}}  \,, \\
\end{aligned}
\label{ansatzf}
\end{equation}
with the phase $\alpha$ is eventually determined by the sign of DMI, and a singularity-free function $\kappa (z\bar{z})$ depends smoothly on the distance from the skyrmion's center.

The equation for $\kappa$ is quite nonlinear and can be solved only numerically. Since $\kappa$ has the dimension of length, we choose to consider a dimensionless function $\widetilde{\kappa}(y)= (\kappa(0))^{-1} \kappa\left(y\, \kappa(0)^{2}  \right)$ with the property $\widetilde{\kappa}(0)=1$. One could then solve the equation for $\tilde{\kappa}(y)$ for different boundary conditions. Our primary interest is to find $\tilde{\kappa}(y)$ on a disc of finite radius which mimics the case of SkX where one skyrmion is surrounded by its neighbors. The pressure exerted by this type of environment is modeled by changing the size of a disc. We found that the function $\tilde{\kappa}(y)$ is nearly invariant against changes of disc radius, in contrast to the value of the dimensionless residue, $\kappa(0)/L$. One hence can model multi-skyrmion configurations by the sum 
\begin{equation}
f(z, \bar{z})= \sum_{j}F\left ( ({\bar z  -\bar z_{j}})/{z_{0}^{(j)}} \right ),
\label{multi-skyrm}
\end{equation}
where
\begin{equation}
F\left (\frac{\bar z}{z_{0}} \right ) \equiv
\frac{z_{0}}{\bar z} \tilde \kappa_{\infty}\left ( \left|\frac{\bar z}{z_{0}}\right |^{2} \right)  \,,
 \label{eq:skyform1}
\end{equation}
with $\kappa_{\infty}$ is the solution on the disc of infinite radius, and $|z_0|$ in this formula is the skyrmion's size. 
We remind that the formula \eqref{multi-skyrm} with arbitrary 
$z_{j}$, $z_{0}^{(j)}$ provided an exact (metastable) solution to \eqref{energy} at $D=B=0$.  In that case skyrmions did not interact, and the energy was given by $ \mathcal{E} = \sum _j \mathcal{E}[f_{j}]$ with the energy of individual skyrmions (chemical potential) $ \mathcal{E}[f_{j}] \equiv 4\pi C $. 
Both DMI and the magnetic field bring characteristic scales into the model, that results in the interaction between skyrmions, which is the main difference between the model (\ref{energy}) and BP model. 


We perform the exact numerical calculation of the energy density for SkX modelled by \eqref{multi-skyrm}  with the use of formulas  (\ref{energy})--\eqref{eq:skyform1}, for 
 the most interesting case of densely packed SkX
by putting $\bar{z}_{j}$ onto triangular lattice. The energy density calculated within the (hexagonal) unit cell of such SkX 
 has two parameters: the unit cell parameter or period of the lattice, $a$, and the radius of the skyrmion, $|z_0|$ ;    one should minimize the density, $\rho = 2/\sqrt{3} E_{cell}(z_0,a)/a^{2}$ over $a$ and $|z_0|$.
We present the results of this minimization below, in Fig.\ \ref{fig:dens} and compare it with other model configurations.

Earlier we showed \cite{Timofeev2019207} that the energy of configuration \eqref{multi-skyrm} can be regarded as the sum of (i) the energies of individual skyrmions, $\mathcal{E}[f_{j}]$, (ii) the pairwise (repulsive) interactions between them, $U_{2}(z_{0},a) =\mathcal{E} \left [  f_{1} + f_{2}  \right] - \mathcal{E}[f_{1} ]  - \mathcal{E}[f_{2} ]$, and (iii) the remaining part, $U_{3}$, which does not fit to these two categories. Because of strong non-linear effects of the model,    $U_{3}$
 turns out to be sizeable (and attractive) and 
largely corresponds to the triple interaction between the nearest skyrmions. Interestingly, as first noticed in \cite{Timofeev2019207}, the exact calculation of the optimal energy per unit cell corresponds with the   
overall accuracy $10^{-3}$ to the approximate expression  
 \begin{equation}
 E_{cell}(z_0,a) =  \mathcal{E}[f_{1}] +3 U_{2}(z_0,a)+   U_{3}(z_0,a). \label{SkXdens} \end{equation}
which takes into account only pairwise and triple interactions between nearest neighbors on the triangular lattice. This surprisingly good approximation means a possibility to discard the contribution from the next-to-nearest neighbors (NNN). It cannot be fully explained in terms of  relative smallness of the pairwise NNN interaction, but rather as a combined effect with the triple NNN interaction of opposite sign. 
This idea is supported by the observation, that the calculated energy in the hexagonal cell around a skyrmion with its six neighbors is nearly identical to the energy, calculated for this configuration with six added NNN skyrmions.

\section{Single helix}
\label{sec:SingleH}

The well known expression  \cite{kaplan1961}  for single helix configuration in magnets with DMI is given by:
\begin{equation}
\mathbf{S} = \hat{c} \cos{\alpha} + (\hat{b} \cos{\left(\mathbf{k} \mathbf{R} + \beta\right)} + \hat{a}\sin{\left(\mathbf{k} \mathbf{R} + \beta\right)}) \sin{\alpha}\,,
\label{skaplan}
\end{equation}
where $\hat{a}, \hat{b}, \hat{c}$ are unit vectors with $\hat{a}=\hat{b}\times\hat{c}$, $\mathbf{k}$ is the helix propagation vector and $\alpha$ is the cone angle. Eq.\ (\ref{skaplan}) is the starting point for analysis of all helical states: conical, cycloidal, etc. The main question of such an analysis is the choice of $\hat{\mathbf{k}}$, $\hat{a}, \hat{b}, \hat{c}$, the values of $k$ and $\theta$. All these parameters are determined by particular form of the Hamiltonian, crystal symmetries, etc.

We are interested in the 2D spatial case, so   $\mathbf{k}$ lies in a plane. Parametrizing the basis as $\hat{a}=(-\sin{\varphi},\cos{\varphi},0)$, $\hat{b}=(-\cos{\theta}\cos{\varphi},\cos{\theta}\sin{\varphi},\sin{\theta})$ and $\hat{c}=(\cos{\varphi}\sin{\theta},\sin{\varphi}\sin{\theta},\cos{\theta})$, one can show 
for arbitrary DMI,  that in the 2D case the vector $\hat{c}$ lies in a plane, $\theta=\pi/2$,  and the cone angle collapses, $\alpha=\pi/2$. It means that the spin configuration becomes 
\begin{equation}
\mathbf{S}_{\varphi}=\left(\begin{array}{crl}
\sin{\varphi} \sin{\left(\mathbf{k}_{\varphi}\mathbf{R} \right)}\\
-\cos{\varphi} \sin{\left(\mathbf{k}_{\varphi}\mathbf{R}  \right)}\\
\cos{\left(\mathbf{k}_{\varphi}\mathbf{R} \right)}
\end{array}\right).
\label{shelix}
\end{equation} 
The angle $\varphi$  defines the plane of magnetization rotation  and in turn determines the direction of $\mathbf{k} _{\varphi}$ for particular form of DMI. 
In this paper we use the relation 
\[ \mathbf{k} _{\varphi} = k (\cos{\varphi},\sin{\varphi},0) \,,\] 
appropriate for our 2D model \eqref{energy}.  This case is realized in the case  
of cubic symmetry of crystal (B20 compounds for example), where Dzyaloshinskii vector is parallel to bonds. 
 Different types of crystal symmetries could lead to different forms of DMI, and the relation between $\mathbf{k}_{\varphi}$ and $\varphi$ could be different.

Actually in the presence of an external magnetic field perpendicular to the plane, Eq.\ \eqref{skaplan} is not an exact solution of the model \cite{dzyaloshinskii1964}. The well-known fact is that in uni-axial magnets with DMI the simple helix also transforms to the chiral soliton lattice (CSL)  \cite{izyumov1984modulated}. 
If spins are modulated in $\hat{x}$ direction and lie in the perpendicular plane, $\mathbf{S} = (0,-\sin{\left(\phi(x)\right)},\cos{\left(\phi(x)\right)})$, then the energy \eqref{dimenergy} takes the form
\begin{equation}
\mathcal{E} = \frac{1}{2}(\partial_{x} \phi(x))^2 - \partial_{x} \phi(x) + b(1-\cos{\left(\phi(x) \right)})\,,
\label{cslen}
\end{equation}
with the resulting Euler-Lagrange equation:
\begin{equation}
\partial^2_{x} \phi(x)=b\sin{\left(\phi(x) \right)}.
\label{sineGordon}
\end{equation}
this is the sine-Gordon equation having the quasi-periodic solution:
\begin{equation}
\phi_0(x) = 2 \ \mathrm{am} \left(\sqrt{\frac{b}{m}} x +\beta \bigg| m \right) + \pi\, , 
\label{cslsol}
\end{equation}
with the elliptic parameter $m$. The minimization of the energy \eqref{cslen} links  this parameter to the field :
\[\frac{ E(m)}{\sqrt{m}}  = \frac{\pi}{4 \sqrt{b}},  \]
whereas the energy density at the minimum, $\rho$, and the pitch $k$ are given by expressions \cite{Aristov02}
\begin{equation}
\rho = -2 b\, \frac{1-m}{m}\,,\quad k = \frac{\pi}{K(m)} \sqrt{\frac{b}{m}}  \,, 
\label{eq:rho-k-m}
\end{equation}
here $K(m)$ ($E(m)$) is a complete elliptic integral of the first (second) kind.

Below we expect that the 1D solution \eqref{cslsol} may be convenient for parametrization of the trial function, although the conditions  \eqref{eq:rho-k-m} do not hold.  In this case we can still use \eqref{cslsol}  as a general model form of deformed helix with one control parameter, $m$:
\begin{equation}
\tilde{\mathbf{S}}_{\varphi} =
\begin{pmatrix}
-\sin{\varphi} \,\sin{ \, 2 \mbox{ am}\left(\frac{K(m)}{\pi}(\mathbf{k} _{\varphi} \mathbf{R} +\beta)\Big|m\right) }\\
\cos{\varphi} \,\sin{ \, 2 \mbox{ am}\left(\frac{K(m)}{\pi}(\mathbf{k} _{\varphi} \mathbf{R} +\beta)\Big|m\right)  }\\
-\cos{ \,2 \mbox{ am}\left(\frac{K(m)}{\pi}(\mathbf{k} _{\varphi} \mathbf{R} +\beta)\Big|m\right) }
\end{pmatrix} \,.
\label{shelixdef}
\end{equation}
 This expression is the extension of Eq.\  \eqref{shelix} with the same spatial period, and additional ``degree of ellipticity''. It coincides with Eq.\  \eqref{shelix}  at $m=0$, $\beta=\pi$. 

\section{Triple helix}
\label{sec:TripleH}

In the literature one can find a statement that SkX state can be modeled by the sum of three helices with zero sum of helix propagation vectors. In particular, it was argued \cite{muhlbauer2009skyrmion} that thermal fluctuations stabilize the superposition of three helices at high temperatures in three-dimensional case. Moreover it has been shown in \cite{adams11}  that second Bragg peaks in neutron scattering can be mostly attributed to  the result of double scattering, and they have insignificant intensities in comparison with the first Bragg peaks.

The simple sum of three helices \eqref{shelix}:
\begin{equation}
\mathbf{S}_{3q}=
\mathbf{S}_{ \varphi=0}+\mathbf{S}_{ \varphi=2\pi/3}+\mathbf{S}_{ \varphi=4\pi/3} + S_0 \hat{e}_z
\label{3qsum}
\end{equation}
has a different magnitude from point to point, i.e. $|\mathbf{S}_{3q}(\mathbf{R})|\neq const$. For the A-phase of 3D compounds, the possibility of this variation can be explained by a closeness to the critical point where the magnitude of magnetization could vary significantly. But in the planar case of our interest at $T=0$, one should expect the constraint $|\mathbf{S}|=1$. Below we consider two ways to obtain the normalized triple helix configuration.

\subsection{Triple helix in the stereographic projection method}

As discussed above, the stereographic projection automatically provides the low-temperature normalization constraint $|\mathbf{S}|=1$, which is convenient for a discussion of  multi-skyrmion configurations. It is tempting to use the method also for construction of a multiple-helix configuration.

One can easily verify that the single helix \eqref{shelixdef} is represented by the function:
\begin{equation}
f_{\varphi}=i e^{i \varphi} \cot \mbox{ am}\left(\frac{K(m)}{\pi}(\mathbf{k} _{\varphi} \mathbf{R} +\beta)  \bigg|m \right).
\label{stereohelix}
\end{equation}
This function has a striped structure of zeros and poles lines. The sum of three helices of the form (\ref{stereohelix}) with different $\mathbf{k}_{\varphi}$
, obeying the relation $\mathbf{k}_{\varphi_1} + \mathbf{k}_{\varphi_2}  + \mathbf{k}_{\varphi_3} = 0$, might appear to be a good choice for description of two-dimensional hexagonal lattice of skyrmions.
However, comparing with the previous formula \eqref{multi-skyrm}, which has simple poles at the centers of skyrmions, we choose a different representation in the form 
\begin{equation}
f_{3q} =\left(\frac1{ {f}_{ \varphi=0}}  + \frac1{{f}_{ \varphi=2\pi/3} } +  \frac1{{f}_{ \varphi=4\pi/3} } \right)^{-1} \,.
\label{3q:ster}
\end{equation}
Parameter $m$ now defines the shape of  skyrmions, and $k$ determines the cell parameter of skyrmion lattice.

\begin{figure}[t]
\center{\includegraphics[width=0.99\linewidth]{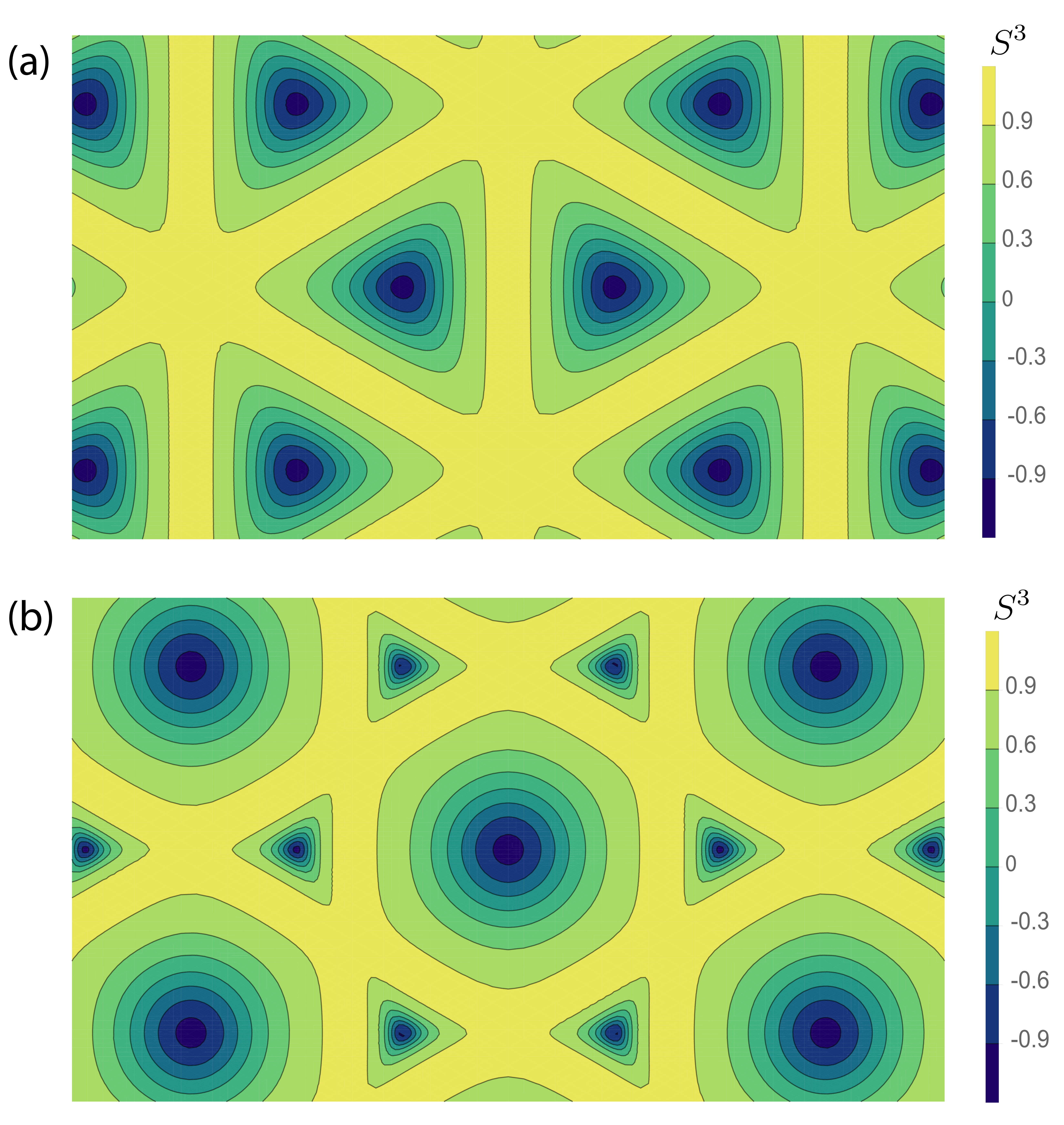}}
\caption{Schematic pictures of two different types of triple helix structures modeled in  stereographic approach: (a) honeycomb lattice appearing for $\beta=\pi$   and (b) kagom\'e lattice for $\beta=0$}
\label{fig:latt}
\end{figure}

In contrast to the combination of 
two functions, e.g.,  $f_{2q} = (( {f}_{ \varphi=0})^{-1} + ( {f}_{ \varphi=2\pi/3})^{-1}  )^{-1} $, where an arbitrary value of $\beta$ could be effectively put to zero by an appropriate shift of the origin, 
the addition of the third helix in \eqref{3q:ster} makes the choice of $\beta$ not harmless. As can be seen in Fig.\ \ref{fig:latt}, two different configurations of  lines of zeros appear,  depending on $\beta=\pi$ or $\beta=0$, corresponding to different topological charge $Q$ per (rhombic) unit cell: for the honeycomb case with $Q=2$, and for the kagom\'e case with $Q=3$.

Our calculation shows  that such a construction of the triple helix leads to the higher energy density as one can see from Fig.\ref{fig:dens}, as compared both to  the SkX ansatz from Sec.\ \ref{sec:SkX} and to the variant of triple helix  considered in the next subsection. 
One can argue that the honeycomb lattice configuration in 
Fig.\ \ref{fig:latt}a  is not tightly packed, which enhances the inter-skyrmion interaction contribution.  
Such argument does not hold for the kagom\'e  configuration in Fig.\ \ref{fig:latt}b, where   centers of skyrmions form a triangular lattice, similarly to our above ansatz  \eqref{multi-skyrm}. 
Our calculation shows that the energy minimization leads to very close estimates in density of topological charge and of Zeeman energy contribution both for \eqref{3q:ster} and  \eqref{multi-skyrm}, whereas the sum of exchange and DMI energy terms is significantly higher in case of \eqref{3q:ster}. The latter observation may be associated
primarily with the inappropriate size of individual skyrmions in the kagom\'e  configuration, since two out of three skyrmions in the unit cell  appear too small  for any elliptic index $m$. 


\subsection{Normalized sum of three deformed helices}

As discussed in Sec.\ \ref{sec:SingleH}, a magnetic field deforms a helix configuration into the more optimal configuration, called as a deformed helix or chiral soliton lattice, Eq.\ \eqref{shelixdef}. 
It seems then only natural to use a more general combination of three such deformed helices \eqref{shelixdef}, instead of simple expression \eqref{3qsum}. To be able to compare the energies of different configurations, we should normalize the resulting magnetization :
\begin{equation}
\begin{aligned}
\tilde{\mathbf{S}}_{3q}= \frac{S_0 \hat{e}_3+\tilde{\mathbf{S}} _{\varphi=0} + \tilde{\mathbf{S}}  _{\varphi=2\pi/3} + \tilde{\mathbf{S}}  _{\varphi=4\pi/3}}{\bigg{|}S_0 \hat{e}_3+\tilde{\mathbf{S}} _{\varphi=0} + \tilde{\mathbf{S}}  _{\varphi=2\pi/3} + \tilde{\mathbf{S}}  _{\varphi=4\pi/3}\bigg{|}}.
\end{aligned}
\label{3qnormDef}
\end{equation}
We call this expression (taken at $\beta=0$) the deformed triple helix (DTH) below.
 
The expression \eqref{3qnormDef} has three variational parameters for energy minimization: a pitch of helices, $k$, the elliptical parameter, $m$, and the additional magnetization perpendicular to the plane $S_0$.  In terms of the resulting SkX structure, the pitch $k$ defines the cell parameter of SkX, while both $m$ and $S_0$ determine the radius and shape of individual skyrmions.
Some analysis shows that $S_0$ takes positive values, and it is the major parameter defining (and reducing) the size of skyrmions. The role of $m$ is   only to adjust the shape of the configuration  \eqref{3qnormDef} ; in contrast to single helix \eqref{eq:rho-k-m} with $m\simeq 1$,  the energy minimization by DTH ansatz \eqref{3qnormDef} yields negative values of $m$ in the whole range of $b$. 
This might be the reason that the energy difference, $\delta\rho$,  between configuration  \eqref{3qnormDef} with $m=0$ and the one with optimal value of $m<0$ is not significant, it is  $\delta\rho \approx 0.005$  at  smaller $b\simeq 0.3$ while $\delta\rho$ tends to zero near $b\simeq 0.75$. 



\begin{figure}[t]
\includegraphics[width=0.99\linewidth]{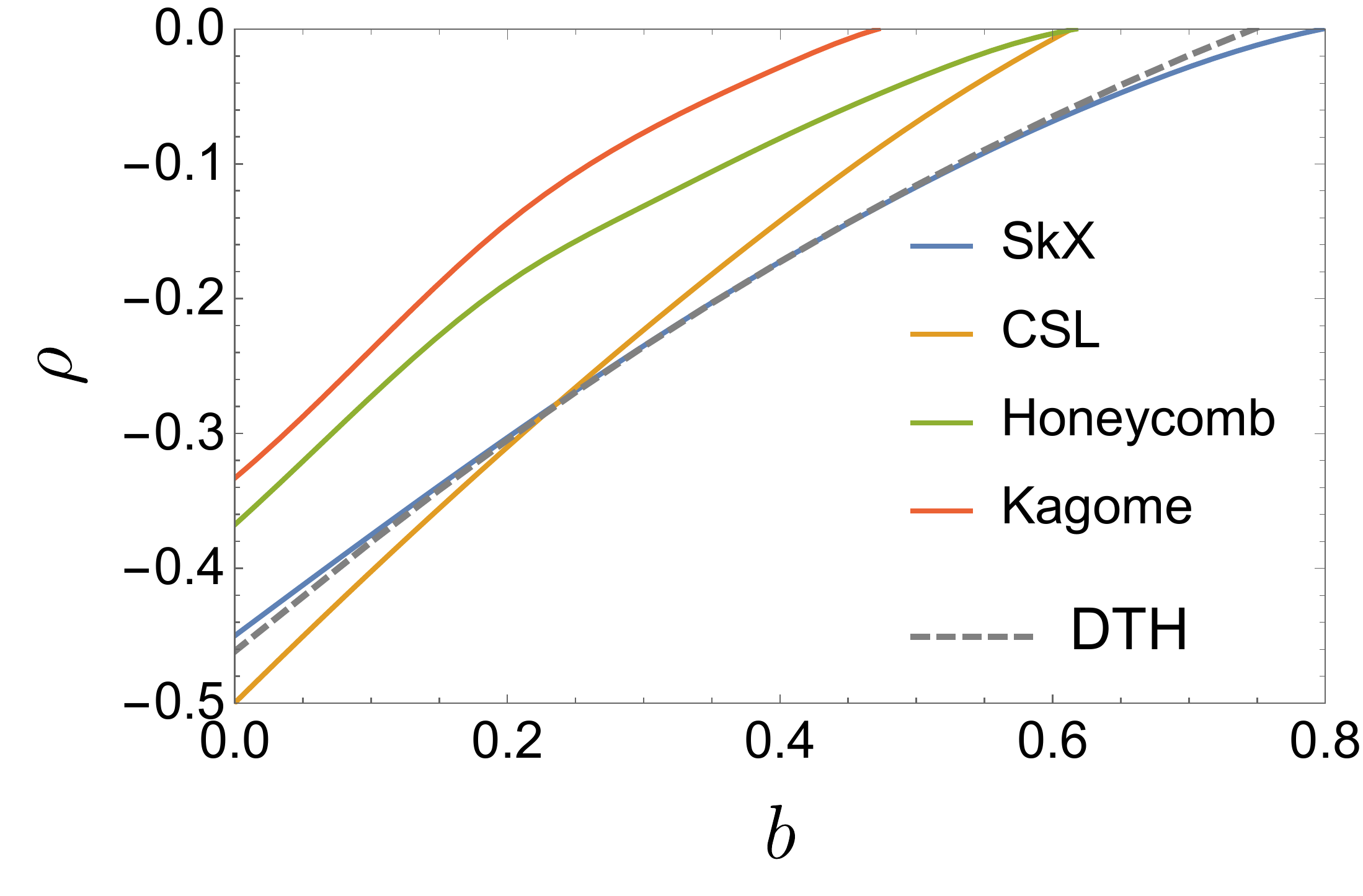}
\caption{Density energy for the Hamiltonian \eqref{dimenergy} and different spin configurations}
\label{fig:dens}
\end{figure}

The energy density found for such an optimal configuration from Eq.\eqref{dimenergy} is plotted as a function of magnetic field in Fig.~\ref{fig:dens}. 
In this Figure we show also the  energy found for SkX ansatz \eqref{multi-skyrm}   and for the single deformed helix \eqref{shelixdef}  with optimal parameters. 
It is seen that at a low external magnetic field $b_{cr1}\alt 0.25$ CSL configuration \eqref{shelixdef}  is energetically favorable, and SkX is advantageous in the intermediate region $b\in (0.25, 0.8)$. In its turn, SkX is destroyed by a magnetic field  at $b_{cr2}\approx0.8$,  when the uniform configuration delivers the energy minimum. This calculations is in a good agreement with previous works \cite{bogdanov1989thermodynamically,BOGDANOV1994255}.
Two above variants of triple helix in stereographic projection (honeycomb and kagom\'e) are shown to be higher in energy.

\section{Comparison of the models} 
\label{sec:comparison}

\begin{figure}[t]
\includegraphics[width=0.99\linewidth]{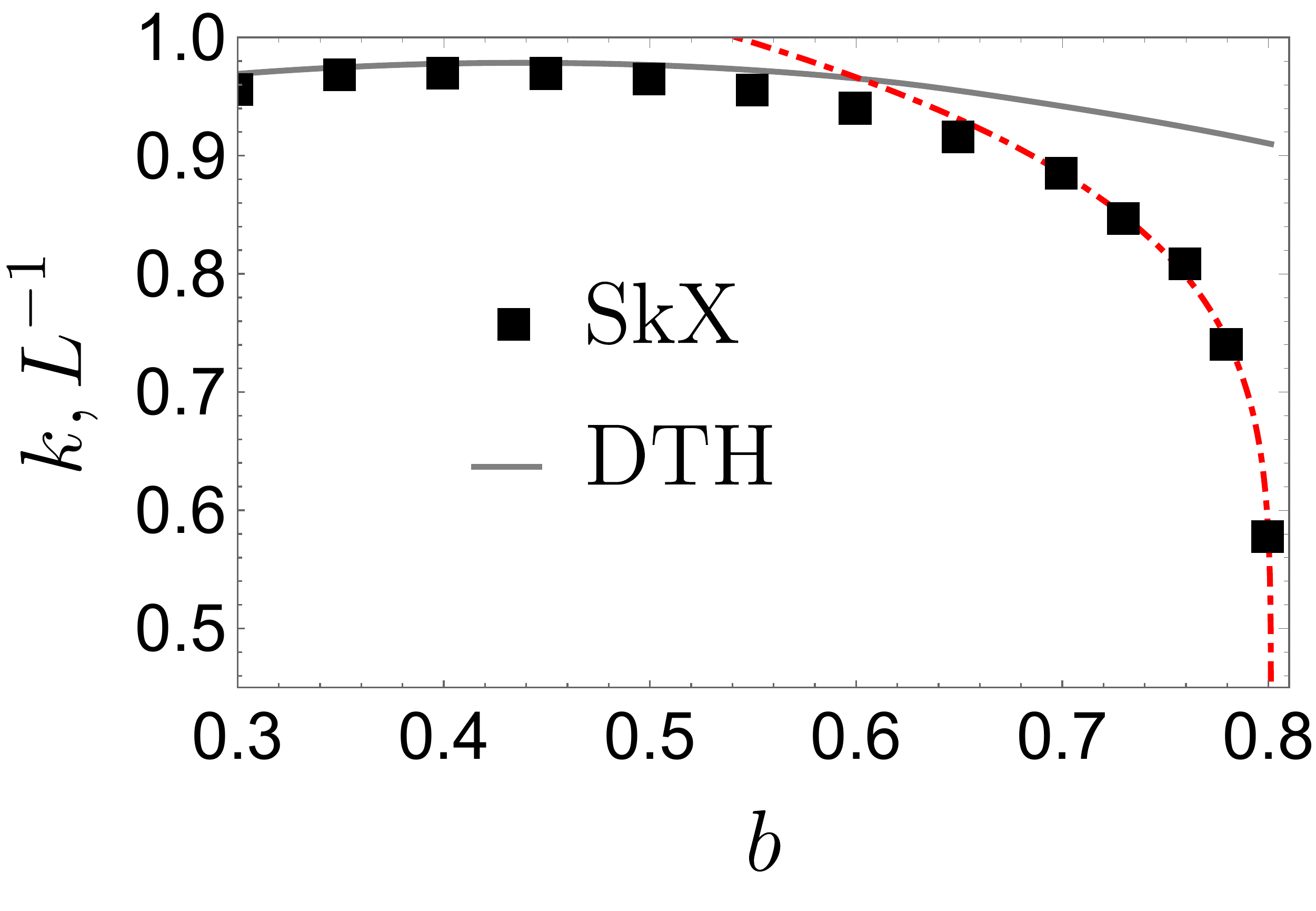}
\caption{Optimal value of modulation vector for ``Triple helix'' and SkX for different values of $b$.
The red dashed line is the fit of SkX values of $k$ as described in text. 
}
\label{fig:pitch}
\end{figure}

We observe  in Fig.~\ref{fig:dens} that the difference in two descriptions, in terms of SkX and deformed triple helix, becomes essential in the region of relatively strong magnetic fields.
More details can be found in analysis of the optimal modulation vector for SkX and DTH, corresponding to inverse unit cell parameter of SkX, $(4\pi/a\sqrt{3})$, and the pitch, $k$, respectively.  
The results are presented in Fig.~\ref{fig:dens}, it is seen that  the DTH solution becomes increasingly different from SkX in the region of high magnetic fields, $b\in (0.6,0.8)$.
In  this region, the SkX with increasing unit cell parameter is eventually described as a rarified gas of weakly interacting skyrmions, and a dissolution or melting of SkX happens at the critical field $b=b_{c2}$. At the same time, the DTH model predicts nearly the same value of helical pitch up  to $b\simeq 0.73$ when   the uniform ferromagnetic  (FM) state  becomes lower in energy.
Considering the density of topological charge $p = k^2 \sqrt{3}/8\pi^2$ as an order parameter in the skyrmion phase, one can say that the  transition to the FM state in the DTH model corresponds to $p$ abruptly changing to zero.
It is instructive to compare this conclusion with SkX ansatz \eqref{multi-skyrm}, where the energy of two skyrmions placed at the distance $R$ from each other behaves \cite{Timofeev2019207} as $E_2 \simeq 2x + A \exp(-R/\ell)$, with $x\sim b-b_{c2}$, correlation length in the FM state $\ell = b^{-1/2}$ and $A\sim1$. Minimization of the energy density, $\sim (x +3A \exp(-R/\ell) )/R^2$ with respect to $R$ leads to $\rho$ depicted in Fig.\ \ref{fig:dens}. It also leads to the dependence of topological charge 
$p\sim (\ell \ln(A/|x|))^{-2}$ and the pitch $k\sim (\ell \ln(A/|x|))^{-1} $ in the vicinity of $b=b_{c2}$.
We show the fit by the latter dependence in Fig. \ref{fig:pitch} by the red dashed line.  
The dependence of $p$ on $b$ near $b_{c2}$ looks qualitatively the same and we do not show it here.

Note that Fig.\ \ref{fig:dens} indicates the 
transitions from SkX phase to helical and FM states at $b_{c1}=0.25$ and $b_{c2}=0.8$, respectively.
According to the recent findings in \cite{huang2020melting}, additional transitions from skyrmion-solid to skyrmion-hexatic and later to skyrmion-liquid phases happen at intermediate fields in thin films of Cu$_2$OSeO$_3$ compound. If we associate the upper critical field found in \cite{huang2020melting} at low temperatures with $b_{c2}$, then we obtain the values for the additional transitions to be $b=0.54$ and $b=0.64$, respectively. 
Comparing these numbers with our Fig.\ \ref{fig:pitch} we see that deviations between our DTH and SkX description happen at higher fields, which correspond to skyrmion-liquid phase in terms of Ref.\ \cite{huang2020melting}. We saw that SkX modelling \eqref{multi-skyrm} provided a better description  at higher fields in terms of the energy. We point out an additional advantage of this description in the anticipated skyrmion-liquid phase, because the SkX modelling with Eq.\ \eqref{multi-skyrm} does not require a long-range ordering in positions of skyrmions, in contrast to DTH and other regular helical structures. 

\subsection{Elastic cross-section}

The simple formula with a linear combination of three helices  (\ref{3qsum}) contains only six spatial Fourier harmonics, i.e.\ only six peaks in the reciprocal space at $\mathbf{k} _{\varphi}$ with $3 \varphi /\pi = 0, \ldots,5$.  
This is what observed experimentally in high-temperature  A-phase in bulk materials \cite{muhlbauer2009skyrmion,adams11}. But as we discussed above, at low temperature for thin films we should think about normalization of magnetization, and elliptical deformations (\ref{3qnormDef}) should also contain higher harmonics, $\mathbf{k} _{\varphi1}+ \mathbf{k} _{\varphi2}$.

The cross-section of the elastic unpolarized neutron scattering on a magnetic structure is given by \cite{squires2012introduction}:
\begin{equation}
\frac{d\sigma}{d\Omega} \propto \sum\limits_{ij}(\delta^{ij}-\hat{q}^{i}\hat{q}^{j})\langle S^{i}_{\mathbf{q}}\rangle \langle S^{j}_{-\mathbf{q}}\rangle,
\label{cross}
\end{equation} 
with $\langle S^{j}_{\mathbf{q}}\rangle=\int{d\mathbf{r} \, e^{i\mathbf{r}\mathbf{q}}\langle S^{j}(\mathbf{r})\rangle}$.  
For periodic structures, such as  SkX and DTH one can represent the cross-section as a sum over reciprocal lattice vectors:
\begin{equation}
\frac{d\sigma}{d\Omega} \propto C_0 + \sum\limits_{m,n} C_{mn} \delta(\mathbf{q}-m\mathbf{b}_1 - n\mathbf{b}_2),
\label{cross2}
\end{equation}
here  $\mathbf{b}_1 = \mathbf{k} _{\varphi=0}$,
$\mathbf{b}_2 = \mathbf{k} _{\varphi=\pi/3}$ and 
 \begin{equation}
\begin{aligned}
 C_{mn} & =\sum\limits_{ij} \bigg( \delta^{ij} - \frac{(m b^{i}_1 + n b^{i}_2)(m b^{j}_1 + n b^{j}_2)}{|m\mathbf{b}_1 + n\mathbf{b}_2|^2} \bigg) \\ & \times \langle S^{i}_{m\mathbf{b}_1 + n \mathbf{b}_2}\rangle \langle S^{j}_{-m\mathbf{b}_1 - n \mathbf{b}_2}\rangle    \,.
\end{aligned}
\label{cmn}
\end{equation}

\begin{figure}[t]
 \includegraphics[width=0.95\linewidth]{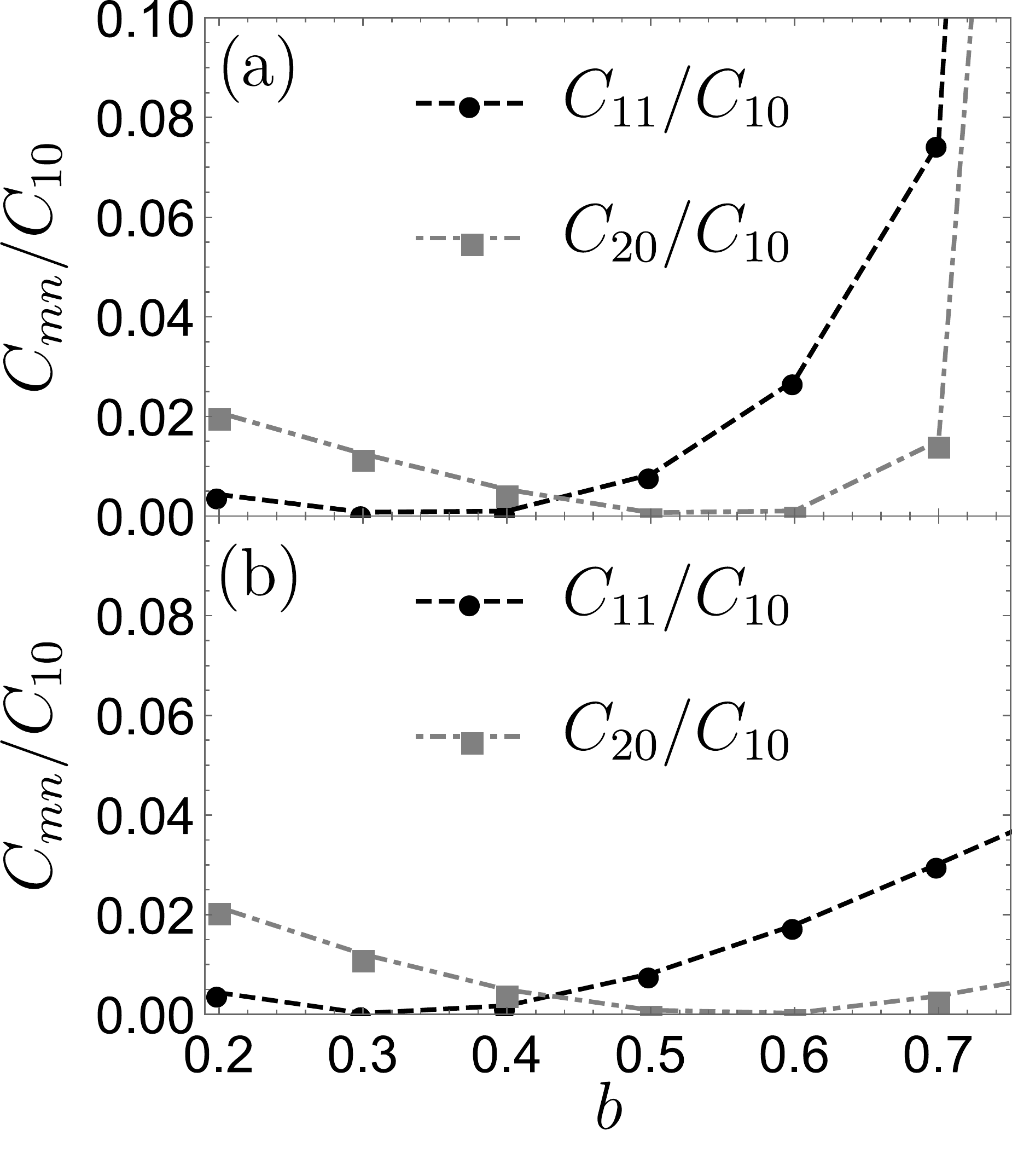}
\caption{  Relative intensities $C_{11}/C_{10}$ and $C_{20}/C_{10}$ for various external magnetic fields, calculated with a) skyrmion crystal (\ref{SkXdens}) approach and b) deformed triple helix (\ref{3qnormDef}), respectively.}
\label{fig:cmnvsb}
\end{figure}

We are interested in  relative values of intensities of higher-order Bragg peaks, $C_{mn}/C_{10}$. 
In our models we find 
that the magnitude $C_{mn}$ rapidly decreases with $m,n$ so that only  $C_{11}/C_{10}$ and $C_{20}/C_{10}$ are  of order of few percents, while the other coefficients are even smaller in the whole range of magnetic field. 
The results of the  calculation for different models of our spin texture are shown in Fig.\ \ref{fig:cmnvsb}. 
It can be seen in this plot that for magnetic fields in the range  $0.3< b< 0.6$,
where SkX and DTH ansatzes yield practically the same energy density,  
both these models give similar results for  $C_{ij}/C_{10}$. This indicates that the spin configuration described by these two approaches is nearly identical.

\begin{figure}[h]
\includegraphics[width=0.95\linewidth]{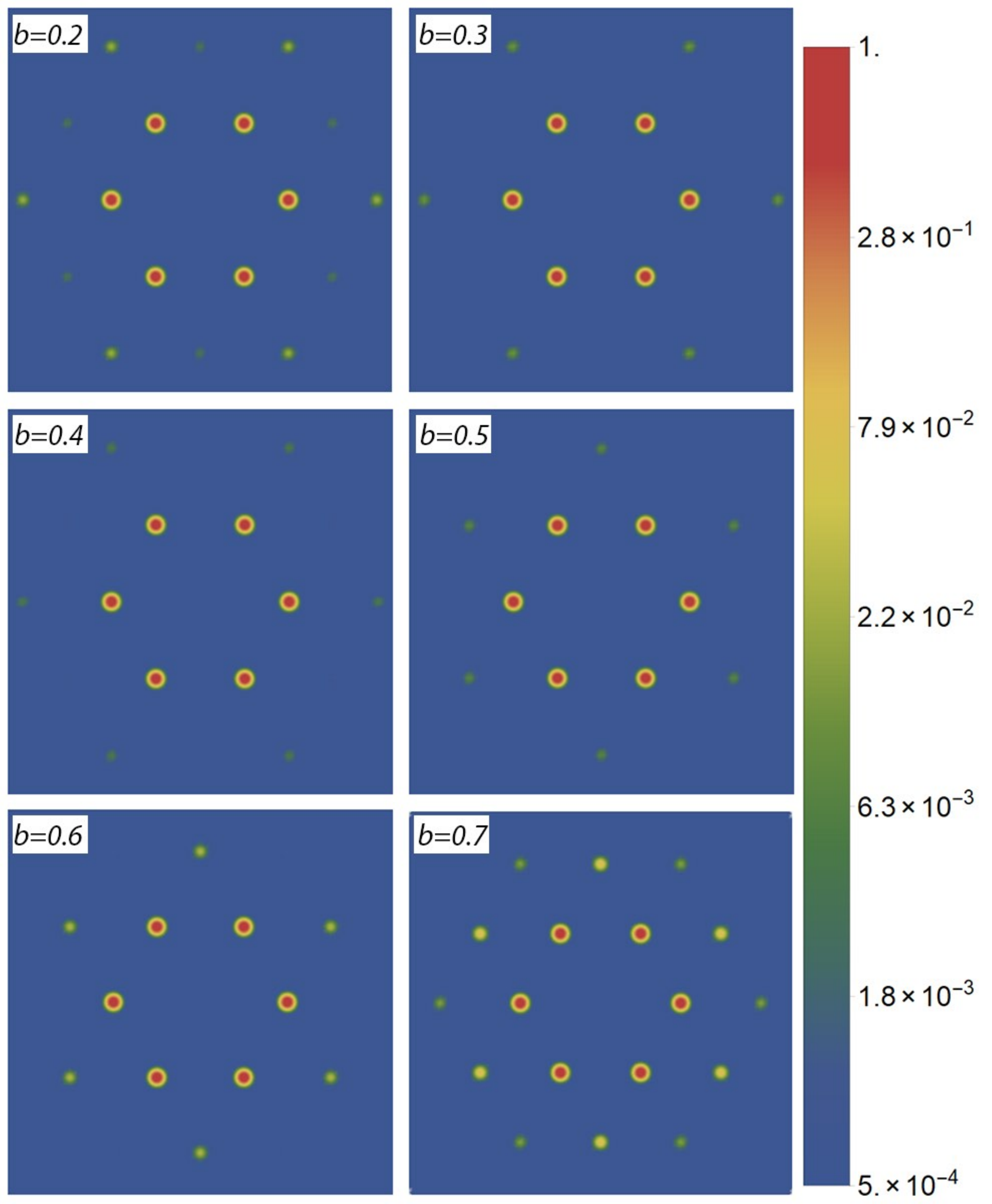}
\caption{Maps of predicted  intensities  for neutron scattering elastic cross-section for different magnetic fields, $b$. Delta functions in \ref{cross2} are approximated by Gaussians. All maps are scaled to the intensity of the first Bragg peaks.}
\label{fig:maps}
\end{figure}

The situation changes in the region of higher magnetic field $0.65 \alt b \alt 0.8$, when   
DTH ansatz fails to reproduce the expected increase in distance between skyrmions.  
We note that for well-separated skyrmions of certain shape within SkX description  \eqref{multi-skyrm} the magnitude of the higher peaks $C_{11}$, $C_{20}$ is defined roughly  by the Fourier image of an individual skyrmion, 
$\langle S^{i}_{\mathbf{q}}\rangle $ taken at $\mathbf{q} = \mathbf{b}_1 +   \mathbf{b}_2$ , $\mathbf{q} = 2\mathbf{b}_1$, respectively.  DTH ansatz, on the contrary, describes somewhat deformed triple helix even at   fields $b\simeq b_{cr2}$, with insignificant admixture of higher harmonics. 
As a result, we see in Fig.\ \ref{fig:cmnvsb} that the values of $C_{11}/C_{10}$ and $C_{20}/C_{10}$ predicted by SkX approach are much larger than for DTH near the melting transition, $b\simeq b_{cr2}$.

According to Refs.\ \cite{seshadri1991hexatic,huang2020melting} (see also \cite{grigoriev2014hexagonal}) the perfect skyrmion crystal is melted before undergoing to uniform ferromagnetic state at $b >  b_{cr2}$.  
Our predictions for the ratio of amplitudes  $C_{ij}/C_{10}$ should partly survive in the intermediate skyrmion liquid phase. Instead of the  well-defined Bragg peaks one observes the concentric circles, corresponding to short range order in the isotropic state. The above intensities $C_{10}$, $C_{11}$, $C_{20}$  should then be associated with the integrated intensities near $|q| = k$, $|q|= k\sqrt{3}$, $|q| = 2k$, respectively.  

At the same time, the above predictions for $C_{11}$, $C_{20}$ cannot be simply compared to Lorentz TEM results, \cite{seshadri1991hexatic,huang2020melting} where the profile of skyrmions has been modelled by $\delta$-function, $\delta(r-r_j)$, as opposed to above Eqs.\ \eqref{stereodef}, \eqref{ansatzf}.

\section{Conclusions}
\label{sec:conclu}

We considered several variants to describe  the skyrmion crystal, formed  at $T=0$ in 2D model of ferromagnet with  Dzyaloshinskii-Moriya interaction and the   magnetic field. 
The first variant is the modification of the stereographic projection method used in the seminal paper \cite{Belavin1975} for the pure $O(3)$ sigma model.
The second approach is the generalization of the triple helix ansatz (\ref{3qnormDef}). The third variant is the combination of the stereographic projection with the triple helix description.

The numerical analysis of the classical energy shows that the first two approaches yield very close estimates  at intermediate values of an external magnetic field, $b$, but are different at lower and higher magnetic fields, close to critical fields characterizing the transitions either to single helix or to uniform ferromagnetic phase. The third approach leads to the  higher energy than the first two approaches in the whole region of $b$. 

Comparing to other methods, the stereographic projection ansatz appears more appropriate at higher magnetic fields, providing lower energy estimate and predicting growing  distance between skyrmions. 

In contrast to the skyrmion A-phase in bulk materials observed at high temperatures, $\sim T_c$, the skyrmion clystal occuring at low temperatures for the 2D case or in layered compounds should lead to  sizeable secondary Bragg peaks. The intensity of these  peaks is non-zero for the saturated local magnetization and depends on details of magnetic structure at low temperatures. Our modeling shows that the intensities of secondary Bragg peaks $C_{11}$ and $C_{20}$ are of order of a few percents of the primary intensity, $C_{10}$. These estimates result from the form factor of individual skyrmions and should apparently survive the melting transition to the skyrmion liquid phase at higher fields.
 

In conclusion, 
analysing the topologically non-trivial ground state of 
the standard model of chiral 2D magnets, we show that its description near the transition to the  ferromagnetic  uniform state is preferable within the stereographic projection method. 
An investigation of the secondary Bragg reflexes in the skyrmion state can give the additional information about the magnetic profile of individual skyrmions.




\begin{acknowledgments} 
We thank  E.V. Altynbaev, S.V. Grigoriev  for useful discussions. The work of V.T. was supported by the Foundation for the Advancement of Theoretical Physics BASIS.  
The work of D.A. was funded in part  by the Russian Foundation for Basic Research (grant No. 20-52-12019) – Deutsche Forschungsgemeinschaft (grant No. SCHM 1031/12-1) cooperation.

\end{acknowledgments}

\end{document}